\documentclass[12pt]{iopart}

\usepackage{epsfig,color,colordvi, ulem}

\usepackage{iopams}

\usepackage{subfigure}

\begin{document}

\newcommand{\up}{{\mid \uparrow \rangle}}
\newcommand{\down}{{\mid \downarrow \rangle}}

\title{Nonuniversality and strongly interacting two-level systems in glasses at low temperatures}

\author{M. Schechter$^1$, P. Nalbach$^{2}$, and A. L. Burin$^3$}
\address{$^1$Department of Physics, Ben Gurion University of the Negev, Beer Sheva
84105, Israel\\
$^2$Westf\"alische Hochschule, M\"unsterstr. 265, 46397 Bocholt, Germany\\
$^3$Department of Chemistry, Tulane University, New Orleans, LA, 70118, USA}

\date{today}

\begin{abstract}

Recent experimental results showing untypical nonlinear absorption and marked deviations from well known universality in the low temperature acoustic and dielectric losses in amorphous solids prove the need for improving the understanding of the nature of two-level systems (TLSs) in these materials. Here we suggest the study of TLSs focused on their properties which are nonuniversal. Our theoretical analysis shows that the standard tunneling model and the recently suggested Two-TLS model provide markedly different predictions for the experimental outcome of these studies. Our results may be directly tested in disordered lattices, e.g KBr:CN, where there is ample theoretical support for the validity of the Two-TLS model, as well as in amorphous solids. Verification of our results in the latter will significantly enhance understanding of the nature of TLSs in amorphous solids, and the ability to manipulate them and reduce their destructive effect in various cutting edge applications including superconducting qubits.
\end{abstract}

\maketitle

\section{Introduction}

Amorphous solids, and many disordered lattices, show remarkable universality in their thermodynamic, dielectric and acoustic properties at low temperatures. For example, their specific heat is roughly linear in temperature, their thermal conductivity behaves roughly as $T^2$, and the internal friction is constant, independent of temperature and phonon wavelength \cite{ZP71,HR86,PLT02}. It is quite striking that the above properties are not only distinct from the behavior of ordered lattices, but that they are quantitatively similar, within a factor of $3$, in materials ranging between very different amorphous solids, disordered polymers and disordered lattices. Furthermore, in all these materials the above characteristics emerge below a rather universal temperature of $T_U \approx 3$K. This striking universality suggests the existence of a general mechanism, which only depends on the disorder itself. This mechanism then dictates the low energy characteristics of disordered systems, and the energy scale of $3$K below which these properties appear. Despite many experimental and theoretical studies throughout the last four decasdes, see e.g. \cite{klein78, sethna85, YL88, solfK94, BNOK98, parshin94, wolynes01, Kuhn03, parshin07, Agarwal13},
this mechanism is not finally understood.

Soon after the discovery of universality, Anderson, Halperin and Varma \cite{AHV72}, and Phillips \cite{Phi72}, suggested that the existence of tunneling two-level systems (TLSs) having a broad distribution of their bias energy $\Delta$ and of their barrier heights, gives rise to the above properties. In what is now referred to as the "Standard Tunneling Model" (STM), the tunneling amplitude of the TLS is defined by $\Delta_0$, and the distribution of TLS parameters is given by $P(\Delta,\Delta_0) \equiv P_0/\Delta_0$. Using the STM to interpret the experimental data, universality reduces to the fact that $l/\lambda \approx 150$ ($l, \lambda$ are the phonon mean free path and wavelength), or that the relevant dimensionless parameter in the theory, the "tunneling strength" $C_0 \equiv P_0 \gamma^2/(\rho v^2)$ is small and universal \cite{YL88}, $C_{0}\approx 10^{-3}$. Here $\gamma$ denotes the TLS-phonon coupling constant, $\rho$ the mass density and $v$ the speed of sound. However, the smallness and universality of the tunneling strength, and the energy scale of $3$K below which universality is observed, are not explained within the STM. Nonuniversal deviations from the STM are observed at lowest temperatures, i.e. $T\lesssim 100$mK, and typically stem from weak interaction between the tunneling systems due to their elastic and electric moments \cite{Burin95,Nalbach04} which are neglected within the STM. Observed magnetic field dependences \cite{Ludwig2002} were attributed to nuclear electric quadrupole moments \cite{Wurger2002,BPFS06,Bartkowiak2013} and paramagnetic impurities \cite{Borisenko2011}. Also, the STM cannot completely account for nonequilibrium and nonlinear dielectric losses in Josephson junction qubits \cite{MPO+10,KWO11,KSG+13,KSB+17}, where the tunneling systems are responsible for the coherence breakdown \cite{SLH+04,MCM+05,KSB+16}.

Recently, within disordered lattice models, it was suggested that TLSs must be separated in two classes, denoted by $S$- and $\tau$-TLSs, based on their local inversion symmetry. Inversion asymmetric ($S$) excitations, {\it e.g.} CN rotations in KBr:CN, interact strongly ($\gamma_{\rm s}$) with the strain, whereas ($\tau$) excitations in which the two local states relate to each other by local inversion
{\it e.g.} CN flips in KBr:CN, interact weakly ($\gamma_{\tau}$) with the strain \cite{SS08,SS09,PS17}. $g \equiv \gamma_{\tau}/\gamma_{\rm s} \approx 0.03$, proportional to the strain in strongly disordered materials and quantifying the degree of deviation from local inversion symmetry, is the small parameter of the Two-TLS model.
All essential aspects of the Two-TLS model, including the symmetry dependent TLS-strain interaction constants and the resulting single TLS density of states (DOS) for the $S$ and $\tau$ TLSs, were thoroughly verified numerically for the disordered lattices \cite{GS11,CGBS13,CBS14,CBS14b}.

Within the Two-TLS model, the weakly interacting ($\tau$) TLSs correspond to the standard TLSs within the STM. However, the Two-TLS model derives a relation between the single TLS DOS of the $\tau$-TLSs and their strain interaction strength, which results in the small and universal value of the tunneling strength. This result suggests that the origin of the low temperature universality, as well as the energy scale of $3$ K determining the universal regime, are consequences of the smallness and universality of deviations from local inversion symmetry in strongly disordered and amorphous materials.
Thus, the analysis, within the Two-TLS model, of the characteristics of the $\tau$-TLSs, suggests a theoretical foundation for results previously obtained by applying the phenomenological STM together with the experimental input of the universality of the tunneling strength.
The strongly interacting ($S$) TLSs are scarce at low energies, and contribute negligibly to the specific heat, internal friction, and thermal conductivity in the universal regime.
However, the predictive power of the Two-TLS model lies in the domination of nonuniversal low-energy properties by the $S$-TLSs, despite their scarcity at the relevant energies.

It is the focus of this paper to obtain and propose within the Two-TLSs model experimental observables which go beyond the STM, and thus provide a clear signature of Two-TLSs physics. This would allow an experimental validation of the Two-TLS model within the disordered lattices, and an experimental test of its relevance for amorphous systems in general. We focus on nonuniversal experimental observables, which depend stronger than quadratically on the TLS-phonon interaction, and can therefore result in a dominant and measurable contribution of the strongly interacting $S$ TLSs.
For these observables we obtain within the Two-TLS model results which
differ markedly with those obtained within the standard tunneling model.

With regard to the relevance of the Two-TLS model to amorphous solids, we note here that while such relevance is not obvious, there are reasons to believe its plausibility (see discussion in Ref.~\cite{SS09}). We note in this respect the strong experimental evidence for the equivalence of the universal phenomena in amorphous systems and disordered lattices \cite{YKMP86,LVP+98,PLC99}, suggesting that a model deriving the apparent universality is applicable to all systems exhibiting the low temperature universal properties. Furthermore, the Two-TLS model was recently proved advantageous in explaining the strain dependence of echo dephasing of TLSs in superconducting qubit circuits \cite{MSSS16}, and the power dependent loss tangent in superconducting microresonators \cite{MPO+10,KWO11,KSG+13,KSB+17}. We note also that nonhomogeneity of the TLS DOS was recently found experimentally in amorphous SiO films \cite{SKW+14},
and that the existence of two types of TLSs was recently demonstrated
in amorphous ${\rm Al_2O_3}$ and ${\rm LaAlO_3}$ films \cite{KSG+13}, where the weakly interacting TLSs were attributed to Hydrogen impurities.

\section{Single TLS DOS in the Two-TLS model}

Within the STM the single TLS DOS is de-facto flat, i.e. $n(E)=P_0\log(E/\Delta_{0}^{{\rm min}})$ for $\Delta_{0}^{{\rm min}}\ll E\ll \Delta_{0}^{{\rm max}}$ and $n(E)=P_0\log(\Delta_{0}^{{\rm max}}/\Delta_{0}^{{\rm min}})$ for $E > \Delta_{0}^{{\rm max}}$. Here $\Delta_{0}^{{\rm min}({\rm max})}$ is the minimal (maximal) tunnel splitting of the TLS, and $\log(\Delta_{0}^{{\rm max}}/\Delta_{0}^{{\rm min}})\approx 10$. Within the Two-TLS model, the STM assumption for the distribution function, $P(\Delta,\Delta_0) \equiv P_0/\Delta_0$, is relaxed. While keeping the distribution of tunnel splittings similar to that of the STM, the Two-TLS model derives the energy dependence $P_0^{\tau}(E), P_0^S(E)$ of the weakly interacting $\tau$-TLSs and the strongly interacting $S$-TLSs, and correspondingly their single TLS DOS, $n_{(\tau/S)}(E)=P_0^{\tau/S}(E)\log({\rm min}[E,\Delta_{0}^{{\rm max}}]/\Delta_{0}^{{\rm min}})$. The resulting single TLS DOS; $n_S(E), n_{\tau}(E)$, differ markedly between the two groups of TLSs.
%
The single TLS DOS $n_{\tau}(E)$ of the $\tau$-TLSs is well described by a Gaussian shape with width $\approx 10$ K, (the energy scale of the maximal S-$\tau$ interaction $J_0^{S\tau}$) except for a dip of relative order $g$ below the energy scale of the maximal $\tau$-$\tau$ interaction, $U_0 = g J_0^{S\tau} \approx 0.1-0.3$ K ~\cite{SS09,CGBS13,CBS14,CBS14b} (true within the practical temperature domain $T > 1$ mK). Although below $3$ K the single $\tau$-TLS DOS resembles the homogeneous DOS assumed for the TLSs by the STM, it differs in the above mentioned dip at low energies \cite{foot2}, and it is markedly different at larger energies, as it quickly diminishes at energies larger than $3$ K.

The single TLS DOS of the $S$-TLSs at relevant energies $E < 10$ K is $\propto E^{\eta}$ \cite{CGBS13}, where for KBr:CN $\eta \approx 1$ \cite{CBS14}.
At finite temperature the gap of the single $S$-TLSs DOS and the dip of the single $\tau$-TLSs DOS are filled, such that $n(E,T) \simeq n(E=T,0)$ for all $E \leq T$.
The ratio $n_S(E)/n_{\tau}(E)$ is energy dependent. The value of $E_{cr}$ for which this ratio equals $g^2$ determines the temperature $T_{cr}$ below which phonon attenuation, being proportional to $n \gamma^2$, is dictated by the $\tau$-TLSs.
Experiments suggest a rather universal temperature range of $T_{cr} \approx 1-3$ K for a wide variety of amorphous solids and disordered lattices. Numerical calculations for KBr:CN find $E_{cr} \approx 1$ K \cite{CBS14}, in agreement with experiment \cite{YKMP86}. This qualitative picture is used below to demonstrate nonuniversal behaviors expected in nonlinear and nonequilibrium properties of interacting TLSs within the Two-TLS model.


\section{Spectral diffusion and nonlinear acoustic absorption}

Experimentally, TLSs can be directly probed by phonon echo experiments \cite{GG76,GSHD79,NFHE04}. However, because of the scarcity of $S$-TLSs \cite{SS09} at energies probed by these experiments so far (of order $10$GHz and below), experiments find only one type of TLSs, with a coupling constant shown to agree with that of $\tau$-TLSs \cite{GS11}. However, under certain conditions the $S$-TLSs can dominate spectral diffusion and the acoustic response.

\begin{figure}[th]
\subfigure[]{
\includegraphics[width=7.5cm]{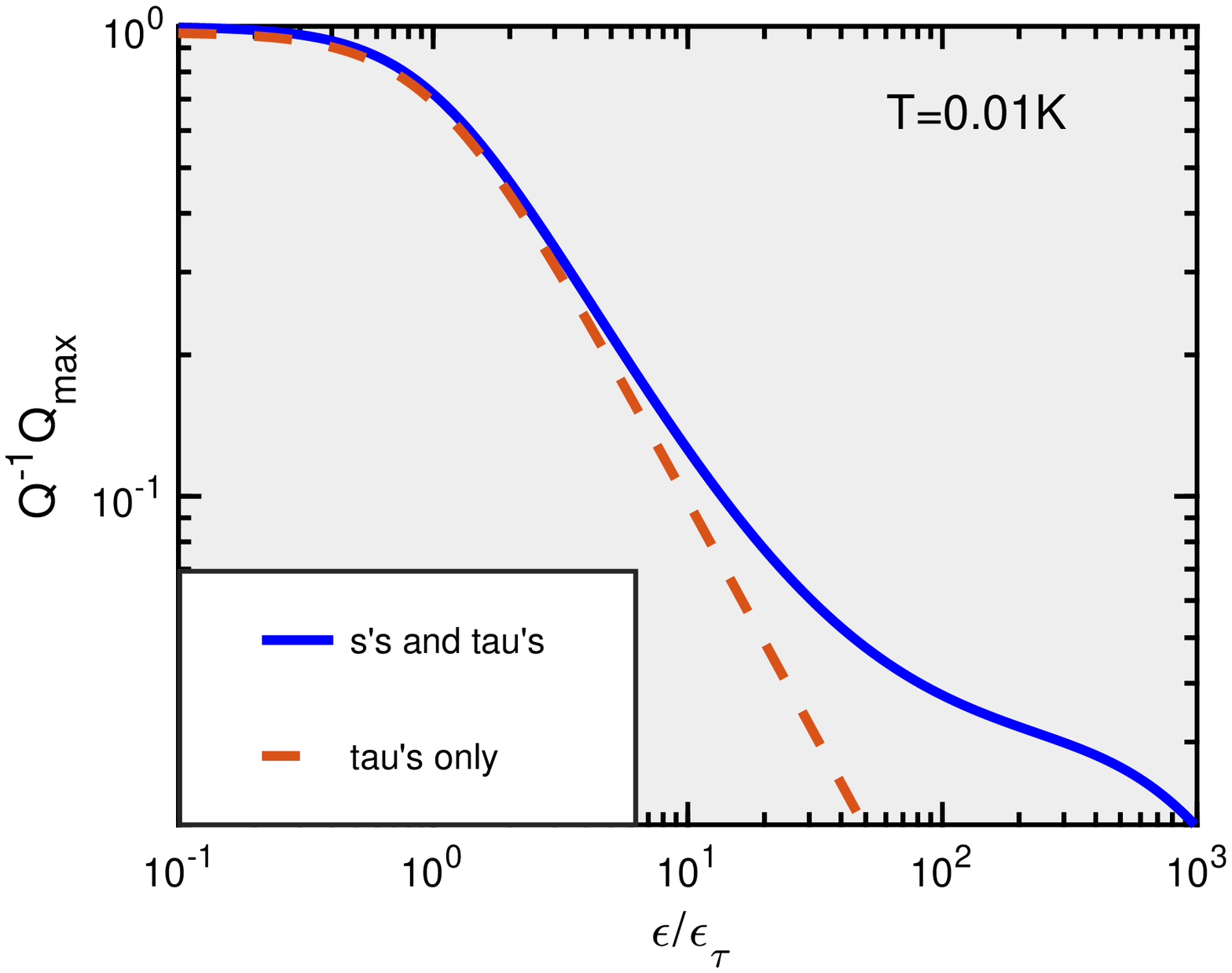}
}
\subfigure[]{
\includegraphics[width=8.0cm]{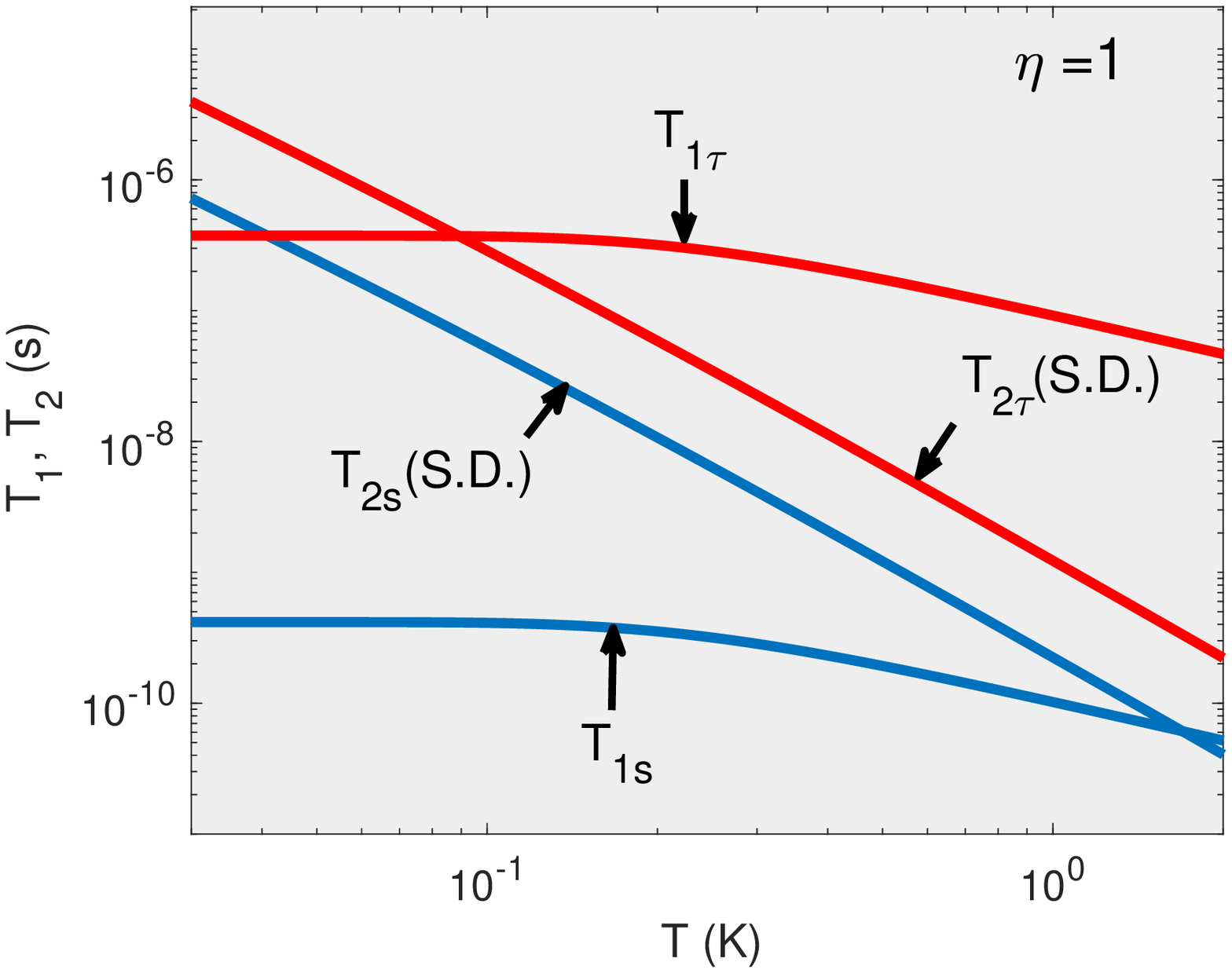}
}
\caption{(a) Relative nonlinear absorption vs. strain at $\nu = 10$ GHz and low temperature $T \ll T_U$. $S$-TLS contribution dominates at high strain because of their smaller relaxation time $T_1$. At $T=0.01$ K the ratio between $\epsilon_c(S)$ and $\epsilon_c(\tau)$ [see equations (\ref{absorbtion}),(\ref{2TLS-absorbtion})] is $1/g$. $T_2$ for the $\tau$-TLSs. At the same time the contribution of the S-TLSs 
(b) Relaxation $T_1$ and spectral diffusion induced phase decoherence $T_2(S.D.)$ times for $\tau$-TLSs and $S$-TLSs. $T_1(S)$ is much shorter than $T_1(\tau)$, as it is proportional to $\gamma^{-2}$, whereas $T_2 \propto \gamma^{-1/2}$. We use the values $g=1/30$ and $\eta=1$ in both plots. }
\label{fig:acoustic}
\end{figure}

\begin{figure}[th]
\begin{center}
\includegraphics[width=7.3cm]{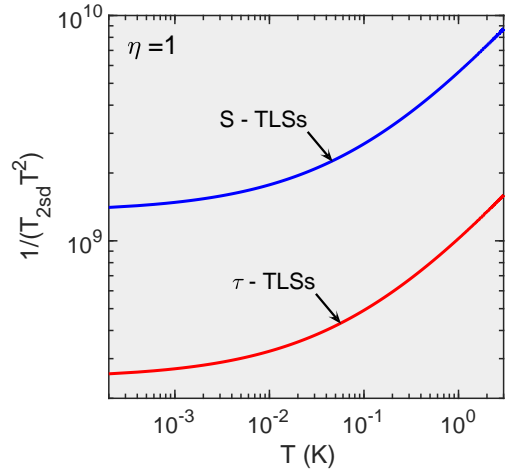}
\end{center}
\caption{Spectral diffusion induced decoherence times for $\tau$-TLSs and $S$-TLSs as function of temperatures, plotted for $g=1/30$ and $\eta=1$. Change of functional form from $T^{-2}$ to $T^{-2-\eta/2}$ is demonstrated. }
\label{fig:spectral}
\end{figure}

At high fields the acoustic response of glasses is nonlinear. For $T \lesssim 0.05K$
the longitudinal and transverse relaxation times are related by $T_2=2T_1$,
and for phonon frequency $\hbar \omega \gg k_{\rm B}T/10$ the acoustic response is given
by the expression \cite{HunklingerArnold}
\begin{equation}
\alpha = \tanh\left(\frac{\hbar\omega}{2k_{B}T}\right) \frac{P_0 \gamma^2}{\sqrt{1+\epsilon^2/\epsilon_c^2}}\frac{\pi \omega}{\rho c^3}
\label{absorbtion}
\end{equation}
where $\epsilon$ denotes strain, and

\begin{equation}
\epsilon_c \equiv \frac{\hbar}{2 \sqrt{2} \gamma T_1} \, .
\end{equation}
Within the Two-TLS model the resonant acoustic absorption is accordingly the sum of the $S$ and $\tau$ contributions, i.e.,
\begin{equation}
\alpha = \tanh\left(\frac{\hbar\omega}{2k_{B}T}\right)
\left[\frac{P_0^S \gamma_{\rm s}^2 }{ \sqrt{1+\epsilon^2/\epsilon_c^2(S)}} + \frac{P_0^{\tau} \gamma_{\tau}^2 }{ \sqrt{1+\epsilon^2/\epsilon_c^2(\tau)}}\right] \frac{\pi \omega}{\rho c^3}.
\label{2TLS-absorbtion}
\end{equation}
Here we assume that the resonant absorption dominates, i.e., that the strain field has a sufficiently high frequency $\nu \sim 10^{10}$s$^{-1}$ \cite{HunklingerArnold}.

At $E,T < 3$K one has $P_0^S(E,T) \gamma_{\rm s}^2 \ll P_0^{\tau}(E,T) \gamma_{\tau}^2$~\cite{SS09}.
This results in the $\tau$ TLSs domination of the universal properties below $T_U$ as mentioned above, and consequently in the acoustic absorption at low intensities. However, at high intensities things are quite different. Considering first low temperatures [$T \lesssim 0.05$ K \cite{BKO13}, see figure \ref{fig:acoustic}(b)], where pure dephasing is negligible, and $T_2 = 2 T_1$, and noting that $T_1 \propto 1/\gamma^2$, we find that
%
$\epsilon_c(\tau) \approx g \epsilon_c(S)$. Thus, for $\epsilon_c(\tau) < \epsilon < \epsilon_c(S)$ acoustic absorption by the $\tau$ TLSs will decrease as $1/\epsilon$. The condition to have an $S$ dominated regime in the acoustic absorption as function of acoustic intensity becomes then [see equation (\ref{2TLS-absorbtion})]
\begin{equation}
P_0^S(E,T) \gamma_{\rm s}^3 > P_0^{\tau}(E,T) \gamma_{\tau}^3 \, .
\label{condition}
\end{equation}
This condition is different from the condition for domination of acoustic scattering in the linear regime, which is dictated by the relative tunneling strengths. It defines a new energy scale $E^* \equiv E_{cr} g^{1/\eta}$ ($E^* \approx 0.05-0.1$ K for $g^{-1} \approx 30-50$ and $\eta \approx 1$) that determines
the frequency $\omega_{low}$ above which $S$-TLSs, despite their small DOS, dominate acoustic absorption at high intensities and low temperatures.
%

The detailed behavior of the acoustic absorption as a function of intensity can be seen in figure \ref{fig:acoustic}(a). The relative resonant absorption at low temperature is shown for $T = 0.01$ K,
where we set $\nu=10$ GHz, $g=1/30$, and $\eta=1$.
It has a marked signature of two plateaus.
The height of the second plateau, as well as its location, are a direct consequence of the $S$ TLS contribution. A clear prediction of the Two-TLS model is the plateau's shift to higher intensity values and its diminished magnitude with the decrease of phonon frequency. Such a measurement as a function of field and frequency
can therefore verify not only the presence of $S$-TLSs, but their DOS as calculated in Ref.\cite{CGBS13}. It should be noticed that the Two-TLS model predicts weakening of the intensity dependence of the absorption similarly to the multiple observations of microwave absorption in Josephson junction qubits and superconducting microresonators \cite{MPO+10,KWO11,KSG+13,KSB+17}. Consistently, the Two-TLS model can possibly account for those observations if one assumes also a difference in the magnitude of dipole moments between the $\tau$ and $S$ TLSs.


Another property which is dominated at $E^* < T < T_U$
by the $S$-TLSs is spectral diffusion. Within the Two-TLS model, the dephasing time associated with spectral diffusion of a $\tau$-TLS is given by \cite{BH77,BLF+13,BMS15}

\begin{equation}
T_{2 \tau}^{SD} \approx 0.1 \left(\frac{k_{\rm B}TP_0^S(0,T)\gamma_{\tau} \gamma_{\rm s}}{\rho c^2 \hbar T_{1S}} + \frac{k_{\rm B}TP_0^{\tau}(0,T)\gamma_{\tau}^2}{\rho c^2 \hbar T_{1\tau}}\right)^{-1/2}
\label{Eq:SD}
\end{equation}
Since $T_{1S} \propto 1/\gamma_{\rm s}^2$ and $T_{1\tau} \propto 1/\gamma_{\tau}^2$, the condition for the first term in the brackets of equation (\ref{Eq:SD}) to dominate in $T_{2 \tau}^{SD}$ is given by equation (\ref{condition}), i.e. the crossover temperature for $S$-TLS domination of spectral diffusion is given by the same energy scale $E^{*}$ noted above for the nonlinear absorption.

The temperature dependence of $P_0^S(0,T) \propto T^{\eta}$ dictates a change in the temperature dependence of $T_{2 \tau}^{SD}$ from being $\propto 1/T^2$ for $T < E^*$ (where the spectral diffusion is dominated by thermal $\tau$ TLSs) to $\propto 1/T^{2+\eta/2}$ for $T > E^*$ (where spectral diffusion is dominated by thermal $S$ TLSs). This is illustrated in figure \ref{fig:spectral} taking $\eta = 1$. Also plotted is $T_2(S)$, which is a factor of $\sqrt{g}$ smaller than $T_2(\tau)$, since a factor of $\gamma_{\rm s}$ replaces $\gamma_{\tau}$ in each term in the brackets of equation (\ref{Eq:SD}). This behavior can be probed in two pulse echo experiments where the echo decays during the decoherence time \cite{BLF+13,LBM+16,MSSS16}.

\section{Nonequilibruim absorption}

The recently developed technique of nonequilibrium loss measurements in the presence of time dependent bias strain field \cite{BKO13} opens another route to detecting the presence of two types of TLSs, strongly and weakly interacting with the strain, along with their respective DOS and its temperature dependence.
Consider the regime where  both the $S$ TLS and the $\tau$ TLS absorptions are saturated with an AC strain field, $\epsilon_{AC}\gamma_{{\rm s},\tau}\gg \hbar / \sqrt{T_{1}T_{2}}$. If one applies a bias strain field $\epsilon_{bias}(t)=v_{bias}t$ to move TLSs away from the ``resonant hole", $\delta E \approx \gamma \epsilon_{AC}$, this move from the hole is much more efficient for the $\tau$ TLSs than for the $S$ TLSs, because of difference in their coupling constants. If the bias is applied fast enough, in the Landau-Zener non-adiabatic regime
$\hbar v_{bias}\gamma_{{\rm s},\tau}\gg \epsilon_{AC}^{2}\gamma_{{\rm s},\tau}^{2}$, then both $S$ TLSs and $\tau$ TLSs move away from the hole, and the absorption approaches its maximum value corresponding to the linear response theoretical limit \cite{BKO13}
\begin{equation}
\alpha \approx \tanh\left(\frac{\hbar\omega}{2k_{B}T}\right)\left[P_0^S \gamma_{\rm s}^2 + P_0^\tau \gamma_{\tau}^2\right]\frac{\pi \omega}{\rho c^3}.
\label{eq:nosaturation}
\end{equation}
At slow bias change, $\hbar v_{bias}\gamma_{{\rm s},\tau}\ll \epsilon_{AC}^{2}\gamma_{{\rm s},\tau}^{2}$, the absorption is linear in the bias

\begin{equation}
\alpha \approx \tanh\left(\frac{\hbar\omega}{2k_{B}T}\right)\left[P_0^S \gamma_{\rm s}^2\frac{\hbar v_{bias}}{\gamma_{\rm s}\epsilon_{AC}^2} + P_0^\tau \gamma_{\tau}^2\frac{\hbar v_{bias}}{\gamma_{\tau}\epsilon_{AC}^2}\right]\frac{\pi \omega}{\rho c^3}.
\label{eq:saturation}
\end{equation}

\begin{figure}[th]
\begin{center}
\includegraphics[width=7.3cm]{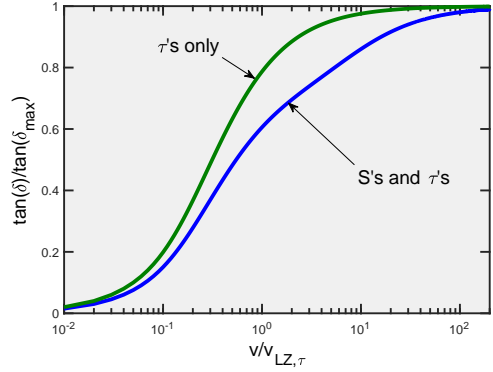}
\end{center}
\vspace{-0.7cm}
\caption{Loss tangent as function of sweep rate for large AC strain field, $\epsilon_{AC}\gamma_{{\rm s},\tau}\gg \hbar / \sqrt{T_{1}T_{2}}$. For intermediate bias rates, $\gamma_{\tau}\epsilon_{AC}^{2}\ll \hbar v_{bias}\ll \gamma_{\rm s}\epsilon_{AC}^{2}$, the change in the loss tangent is dominated by the $S$-TLSs, as the $\tau$-TLSs are nearly fully unsaturated by sweep bias.  }
\label{fig:spectral2}
\end{figure}

However there exists the broad intermediate regime, $\gamma_{\tau}\epsilon_{AC}^{2}\ll \hbar v_{bias}\ll \gamma_{\rm s}\epsilon_{AC}^{2}$, where the field changes too slow to significantly affect the nonlinear absorption by $S$-TLSs while the linear absorption already takes place for the $\tau$-TLSs. In this regime the absorption behaves as (within logarithmic accuracy, see \cite{BKO13,KGS+14})
\begin{equation}
\alpha \approx \tanh\left(\frac{\hbar\omega}{2k_{B}T}\right)\left[P_0^S \gamma_{\rm s}^2\frac{\hbar v_{bias}}{\gamma_{\rm s}\epsilon_{AC}^2} + P_0^\tau \gamma_{\tau}^2\right]\frac{\pi \omega}{\rho c^3}.
\label{eq:Ssaturation}
\end{equation}
Thus, the Two-TLS model predicts the existence of three regimes in the functional dependence of the nonequilibrium absorption as function of the rate of bias change. At the low and high rate regimes, the $\tau$-TLSs dictate the slope of the absorption as function of rate, and the fully unsaturated absorption, respectively, However, in the intermediate regime, where $\tau$-TLSs are fully unsaturated, the $S$-TLSs determine the slope of the absorption as function of bias rate, as is illustrated in figure \ref{fig:spectral2}. This slope is proportional to the DOS of the $S$-TLSs, which can be thus monitored via the tuning of temperature and frequency of the measuring field. We note that we consider here bias fields which are larger than the size of the hole, but smaller than the probed energy. Larger bias fields can drive the $S$-TLSs far from equilibrium, and will be considered elsewhere.

\begin{figure*}[t!]
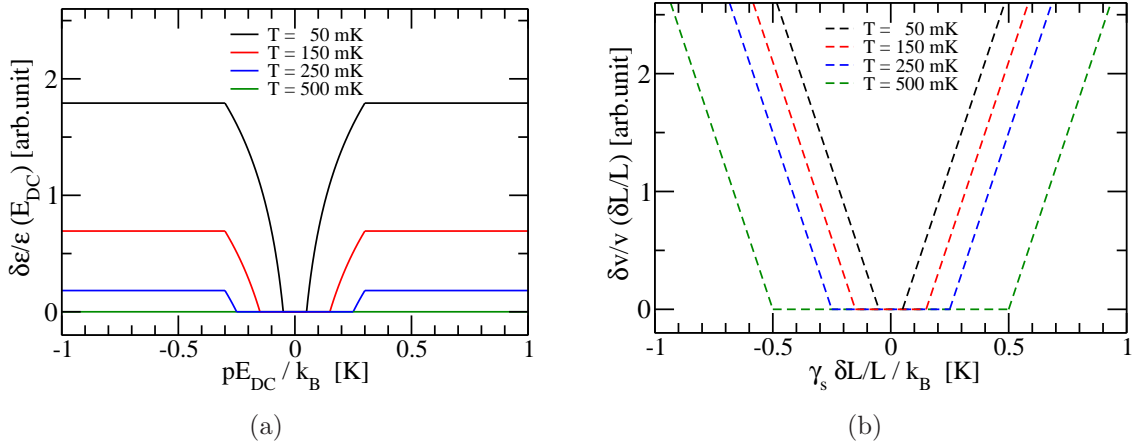

\hspace{5mm}
\subfigure[]{
\includegraphics[width=7cm]{DipolgapPrediction_dielec.eps}
}
\hspace{5mm}
\subfigure[]{
\includegraphics[width=7cm]{DipolgapPrediction_acoust.eps}
}
\caption{ Illustration of (a) dielectric response as function of electric field, and (b) the change in acoustic response as function of strain for various temperatures. The dielectric response saturates at applied bias and temperature which are of the order of $0.3$ K. Neither behavior is observed for the acoustic response, which persists to large bias. We adopted $\eta=1$ here.}
\label{fig:Tad}
\end{figure*}

\section{The dipole gap of the $S$ and $\tau$ TLSs}

Amorphous solids show a dip in their relative dielectric and acoustic responses at low temperatures, as function of an applied external electric or strain field \cite{Burin95,RNO96,NRO96,NRO98,Ludwig03,Nalbach04}. This dip is attributed to a corresponding dip in the DOS of single TLSs at low energies due to many-body effects. Applying external fields allows temporarily to eradicate the dip partially. This causes a change in the TLS DOS $\delta n(E,\Phi_{\rm DC},T) \equiv n(E,\Phi_{\rm DC},T) - n(E,\Phi_{\rm DC}=0,T)$
and for the weakly interacting TLSs, within the STM, one finds for $k_{\rm B} T<\Phi_{\rm DC} < U_0$, \cite{Burin95}
\begin{equation} \delta n(E,\Phi_{\rm DC},T) = \tilde{C}_0 \ln\left( \frac{\Phi_{\rm DC}}{k_{\rm B} T} \right) \cdot n(E,T)
\end{equation}
with $\tilde{C}_0=(\pi/3)C_0\ln(\Delta_0^{\rm max}/\Delta_0^{\rm min})$.
Here $\Phi_{\rm DC}=pE_{\rm DC}$ with dipole moment $p$ of the TLS for an external DC electric field $E_{\rm DC}$; and $\Phi_{\rm DC}=\gamma \epsilon$ for an external DC strain. 
The energy $U_0$ denotes the largest TLS-TLS interaction, above which the dip in the single TLS DOS vanishes.
At the same time a corresponding nonequilibrium change in the dielectric and acoustic response results, and is given
by \cite{Burin95}
\begin{equation}\label{dgsingle}
\delta\chi_{\rm neq}^{(d,a)}(\Phi_{\rm DC},T) = \tilde{C}_0 \ln\left( \frac{\Phi_{\rm DC}}{k_{\rm B} T} \right) \cdot \chi_{\rm eq}^{(d,a)}(T) .
\end{equation}
Herein, $\chi_{\rm eq}^{(d,a)}(T)$ is the equilibrium response of the TLSs, and $k_{\rm B} T \ll \Phi_{\rm DC} < U_0$.
Typically the change in the dielectric constant is studied as dielectric response $\chi^{d}=\delta \epsilon/\epsilon$ and the change in the speed of sound as acoustic response $\chi^{a}=\delta v/v$. Note that the overall magnitude is determined by the square of the electric dipole moment $p$ in the dielectric response $\chi^{d}_{\rm eq}\propto p^2$ and by the square of the TLS-phonon interaction constant $\gamma$ in the acoustic response $\chi^{a}_{\rm eq}\propto \gamma^2$.

Let us analyze the electric and acoustic responses within the Two-TLS model. As a result of the form of the DOS of the $S$-TLSs and $\tau$-TLSs discussed above, one finds a nonequilibrium response
\begin{equation}\label{dgstau}
\delta\chi_{\rm neq}(\Phi_{\rm DC},T) = \delta\chi_{\rm neq}^{(\tau)} \left( \Phi_{\rm DC}^{(\tau)},T \right) + \delta\chi_{\rm neq}^{\rm (S)} \left( \Phi^{\rm (S)}_{\rm DC},T \right) .
\end{equation}
The contribution of the $\tau$-TLSs to the nonequilibrium response is similar to that given by equation (\ref{dgsingle}) at low energies.
Within the Two-TLS model, the maximal $\tau$-TLS - $\tau$-TLS interaction dictating the energy scale $U_0$ above which the dip in the DOS of the $\tau$-TLSs vanishes is calculated for CN flips in KBr:CN to be $U_0\simeq 0.1$K \cite{CBS14b}, and we expect it to be $\approx 0.1-0.3$ K in other materials \cite{SS09,CGBS13}.
Furthermore, within the Two-TLS model, at energies larger than $3$ K the single $\tau$-TLS DOS diminishes abruptly \cite{CGBS13} (again, supported by direct calculations for single CN flips DOS in KBr:CN \cite{CBS14,CBS14b}). We thus expect $\delta\chi_{\rm neq}^{\rm (\tau)}$ to increase logarithmically as function of $\Phi_{\rm DC}$ at low biases, saturate at $\Phi_{\rm DC}\simeq U_0\simeq 0.3$K, and diminish at bias energies larger than $\approx 3$K. The nonequilibrium response $\delta\chi_{\rm neq}^{\rm (S)}$ of the $S$-TLSs is additive on top. The DOS of the $S$-TLSs exhibit a power law gap, i.e., $\propto E^\eta$ up to an energy $E_{\rm max}^{\rm S}\simeq 10$K, which results in
\[
\delta\chi_{\rm neq}^{\rm (S)} \left( T, \Phi^{\rm (S)}_{\rm DC} \right)
= \delta\chi_{0}^{\rm (S)}(T) \cdot \left( \frac{\Phi^{\rm (S)}_{\rm DC}}{k_{\rm B} T} \right)^\eta
\]
in the relevant temperature and energy regime $k_{\rm B} T<\Phi^{\rm (S)}_{\rm DC}<E_{\rm max}^{\rm S}$, since application of the external field fills the gap up to $\Phi^{\rm (S)}_{\rm DC}$. Here $\delta\chi_{0}^{\rm (S)}(T)$ is the $S$-TLS contribution to the response at $\Phi^{\rm (S)}_{\rm DC}=0$, a result of the filling of the $S$-TLS gap at finite temperature.

As both the bias field and the measuring field can be applied electrically or acoustically, four different protocols exist: electric (bias) - electric (measurement) (EE), electric - acoustic (EA), acoustic - electric (AE), and acoustic - acoustic (AA). Here we show that at $\Phi^{(\tau)}_{\rm DC} \lesssim 0.3$ K and $T < \Phi^{(\tau)}_{\rm DC}$, in the EE and AE protocols the contribution of the $\tau$-TLSs dominates. Domination of the EA protocol is $\eta$ dependent, and comparable for $\eta \approx 1$, and the AA protocol is dominated by the contribution of the $S$-TLSs. Let us consider e.g. $\Phi^{(\tau)}_{\rm DC} = 0.1$ K. This results in a similar electric bias for the $S$-TLSs, but in an acoustic bias of $\Phi^{\rm (S)}_{\rm DC} \approx 3$ K, as the latter is proportional to $\gamma$. Considering first the AE protocol, the contribution of the $\tau$-TLSs is $\propto g \chi_{\rm eq}$, in accordance with the size of the dip in the $\tau$-TLS DOS at low energies, $\delta P_0^{\tau} \approx gP_0^{\tau}$. The DOS of the $S$-TLSs at $3$ K is $\approx g^2$ times smaller than the $\tau$-TLS DOS, and accordingly their contribution to the $\delta \chi_{\rm neq}$ is $g$ times smaller than that of the $\tau$-TLSs. The $\tau$-TLSs clearly dominate the EE protocol, as at $0.1$ K $P_0^S \approx (0.1/3)^\eta g \delta P_0^\tau \approx g^{1+\eta} \delta P_0^\tau $. However, for the acoustic response, which is proportional to $\gamma^2$, things are quite different. The latter relation for the DOS of the $\tau$-TLSs and $S$-TLSs at $0.1$ K results in their similar contribution to the acoustic response within the EA protocol for $\eta \approx 1$, where the $S (\tau)$ TLSs dominate the response at $\eta >(<) 1$. However, upon acoustic bias, the $S$-TLSs clearly dominate the acoustic response, by a factor of $\approx 1/g$, as $P_0^S(3 {\rm K}) \approx g \delta P_0^{\tau}(0.1 {\rm K})$, whereas the contribution of each TLS is $\propto \gamma^2$. We note that the EE, AE, and EA protocols have been already performed in experiments (see e.g. in Refs. \cite{RNO96,NRO96,NRO98}). However, to the best of our knowledge, the AA protocol, where the S-TLSs dominate was not measured, nor, naturally, were all four protocols measured in the same system.

The domination of the $S$-TLSs in the acoustic nonequilibrium response within the AA protocol has some marked and measurable consequences, which allows us to predict clear distinctions between the response within the EE and AA protocols: (i) Within the EE protocol the response saturates at $\Phi_{\rm DC} \approx 0.3 $K, whereas no such saturation is predicted for the AA response, as the DOS of the $S$-TLSs grows at all relevant energies. (ii) EE response nearly saturates at $T \approx 0.3$ K, where no such saturation appears in the AA response.

In addition, for a single species of TLSs one clearly gets, for a given $\Phi_{\rm DC}$ the relation $\delta \chi (AA) \delta \chi (EE) \sim \delta \chi (AE) \delta \chi (EA)$. However, within the Two-TLS model, and because of the $S$ domination of the AA response, one obtains $\delta \chi (AA) \delta \chi (EE) \gg \delta \chi (AE) \delta \chi (EA)$.
In figure \ref{fig:Tad}(a) we illustrate the form of the electric response to electric bias (EE protocol), dominated by the $\tau$-TLSs, and in
figure \ref{fig:Tad}(b) we illustrate the form of the acoustic response to acoustic bias, dominated by the $S$-TLSs.

\section{Discussion}

We address, within the Two-TLS model, several acoustic and dielectric properties which depend on the various material properties ($P_0, \gamma, \rho, c$) not simply through the dimensionless tunneling strength. These properties therefore do not show universality at low temperatures. One consequence of the different parametric dependence of these properties is the different relative contribution of the $S$-TLSs, allowing us to give clear predictions to the presentation of these yet unobserved asymmetric TLSs in various phenomena. Experimental protocols testing our predictions and allowing the measurement of the DOS of S and $\tau$ TLSs are suggested. Such experiments, performed both on disordered lattices, for which the validity of the Two-TLS model has been thoroughly demonstrated, and on amorphous solids, will provide answers with regard to the equivalence of the mechanism behind universality in these two systems; with regard to the bounds of universality; and with regard to the nature of the TLSs in both these systems. Such answers may prove useful in enhancing our ability to control TLS behavior, and e.g. limit their decoherence of superconducting qubits and nano-mechanical oscillators.

Our estimates for the various crossover temperature are approximate, and can somewhat vary between materials. Also, we assumed, as is the case in KBr:CN, that the two types of TLSs have a similar electric dipole moment. However, our conclusions for the dielectric and acoustic responses can be generalized to include systems in which the dipole moment of the two types of TLSs differ significantly \cite{KWO11}.

\section*{Acknowledgements}

We would like to thank C. Enss and A. Luck for useful discussions. M. S. acknowledges support from the Israel Science Foundation (Grant No. 821/14) and from the German-Israeli Foundation (GIF Grant No. 1183/2011). A. B. acknowledges support from the NSF Epscore LINK program, award number 555483C1. P. N. and M. S. acknowledge financial support from the Deutsche Forschungsgemeinschaft (project NA 394/2-1).


\begin{thebibliography}{10}
\bibitem{foot2}
This dip is observed in the standard theory when extending the STM with the dipole gap theory \cite{Burin95}.
\end{thebibliography}
\end{document}